\input{aipcheck}
\documentclass[
    ,final            
  ]
  {aipproc}

\layoutstyle{8x11double}

\newcommand\doingARLO[2][]{%
  \ifx\mmref\undefined #1\else #2\fi
}

\begin{document}

\title{Common Envelope Evolution through Planetary Nebula Eyes}

\classification{<Replace this text with PACS numbers; choose from this list:
                \texttt{http://www.aip..org/pacs/index.html}>}

\author{Orsola De Marco}{
  address={American Museum of Natural History}
}

\author{Maxwell Moe}{
  address={University of Colorado}
  ,altaddress={REU student at the American Museum of Natural History} 
}

\begin{abstract}
The common envelope interaction is responsible for evolved close binaries.
Among them are a minority of central stars of planetary nebula (PN). Recent
observational results, however, point to most PN actually being in binary
systems. We therefore ask the question if it is feasible that most, or even
all Galactic PN derive from a common envelope interaction. Our recent
calculation finds that if all single and binary primary stars with mass between $\sim$1-8
M$_\odot$ eject a PN, there would be many more PN in the galaxy 
than observed. On
the other hand, the predicted number of post-common envelope PN is more in
agreement with the total number of PN in the Galaxy. This is a new
indication that binary interactions play a functional role in the creation of PN
and an encouragement to intensify efforts to detect binary companions.
\end{abstract}

\maketitle


\section{Introduction}

Most planetary nebulae (PN) are not spherical, but have elliptical or
bipolar shapes \citep{Sok97}. 
The explanation of why Asymptotic Giant Branch (AGB)
stars would more often than not eject a non spherical PN has been the 
center of a hot debate in the field.
Among the theories of the origin of the asymmetry, are
fast rotation of the AGB star, global magnetic fields and binary 
interactions (see \citet{BF02} for a review).
However, none of these theories allows the case to rest. 
AGB stars are large and are not
known to rotate at rates that would allow them to eject a sufficiently 
equatorially-concentrated wind. AGB stars and post-AGB central stars of PN 
are known to posses
magnetic fields, but these are unlikely to have the global structure 
capable of shaping the AGB wind and induce the observed PN morphologies
\citep{Sok05}.

The large variety of binary parameters allows interactions 
scenarios capable of creating any type of morphology. 
\citet{Sok97} lists
these scenarios, which include binary interactions between the primary 
star and stellar or sub-stellar
companions with a variety of separations: the outcome of the interactions vary
between binaries that do not enter CE, to those that do enter a CE phase and 
emerge from it as a close binary, to those that enter a CE phase and merge.
To each scenario corresponds a specific PN morphology.

The problem with these scenarios is that, appealing as they may
be, it is difficult to constrain them observationally.
Thus far, there is conclusive 
evidence that only about 10\% of all PN harbour close binary
central stars \citep{Bon00}, with an additional 10\% with wide binary 
central stars \citep{Cia+99}. 
These numbers, however, are biased because the close-binary detection 
technique of \citet{Bon00} can only detect binaries with periods shorter 
than a few days. 
In an attempt at determining the exact number of close-binary central stars
\citet{DeM+04} carried out a radial velocity (RV) survey of central stars
of PN in the northern hemisphere and \citet{BA04} have done the same
with a sample in the southern hemisphere [Bond, these proceedings].
Their results indicate that most of the central stars
observed have radial velocity variability. If all the RV-variable stars
are binaries, these results imply that PN are by and large a CE
phenomenon.

Unfortunately, the sampling
of the observations did not allow either study to determine periods (except
in the case of IC4593 for which a period of 5.1 days was convincingly
detected).
With one period and small sample sizes, these results are indicative 
but not conclusive. RV variability could, for instance, be due to wind 
variability for at least half of the samples.
The RV variable stars are the subject of a new 
observing campaign, at echelle resolutions, which will
determine the impact of wind variability and, hopefully, detect
periods (De Marco et al., in prep.). 

In addition to the challenge of accounting for binarity in the central star
population {\it observationally}, there is the additional {\it theoretical} 
challenge of determining what to
expect. Theory can guide observations, but thus far our knowledge of the
interaction that leads to a PN and a short period binary central
star is deficient (for a review see \citep{IL93}). The CE is
thought to be responsible for the production of short period binaries,
but the details of the interaction, e.g., the efficiency with which
the secondary transfers its orbital gravitational energy and angular momentum
to the envelope of the primary, are murky at best.

In this contribution we 
carry out a simple calculation of the
expected number of galactic PN deriving from CE ejections as well as
report on some progress on CE simulations.

\section{Population synthesis model of the galactic PN population}

Detecting binary companions to the bright component of central stars of PN
is an extremely hard task. Radial velocity surveys are time-consuming and
if unknown periods are to be determined, the correct time sampling is key.
This is hard to achieve due to restrictions in telescope scheduling, as
well as bad weather or any other unpredictable factors. This is why,
after four years of successful telescope allocations, the northern sample
of \citet{DeM+04}
still stands at 11 objects with only one period determined.

Successful telescope proposals depend on convincing the time
allocation committees that the
time devoted to the project will produce results. This is particularly
hard when results have been slow coming. In addition,
two objections have repeatedly arisen when the high RV-variability 
results are presented at talks. One of them is why should single post-AGB stars {\it not} eject a PN.
The other is
that the main sequence {\it close} binary population is not
sufficiently large to sustain a galactic PN population which derived mostly
from CE interactions, and therefore single stars
are needed in the production of PN. 

The former objection to the binary PN hypothesis has been investigated by
\citet{SS05} who argue that the spherical PN produced by single AGB stars
could be much less likely to be detected because they are
inherently less luminous.
Below, we attempt to answer the latter objection by constructing a population
synthesis toy model. A more sophisticated version
of the calculation below is underway (Moe \& De Marco, in prep.).

\subsection{The number of PN deriving from common envelope interactions}

\begin{itemize}
\item To start, we assume that only systems that undergo a CE {\it on the AGB}
will result in a PN. This is because very few central stars of PN have ever
been found to be post-Red Giant Branch (RGB)
stars (two examples of post-RGB central stars are EGB5 and PHL932;\citet{Men+88}). We 
therefore exclude systems that undergo a CE on the
RGB, i.e., those main sequence binaries with separations smaller than
$\sim$100~R$_\odot$ \citep{Dor+93}.

\item We do not account for systems that have undergone two AGB CEs, i.e.,
both the primary {\it and} secondary have evolved through the AGB expansion
and each time a CE has ensued. These systems are responsible for WD-WD
binary central stars. We will deal with these interesting systems in our refined
calculation but forecast that they are infrequent and will not add much to the
total number of PN with close binary central stars.

\item {\it Determination of the total number of binary systems in the Galaxy.}
First, we summed the matter contributed by stars
in the galactic disk, bulge, and halo (7.5$\times$10$^{10}$~M$_\odot$;
\citet{DB98,Mer+98}).  
The total number of stars in 
the galaxy is then calculated by normalizing the luminous matter to the 
average mass of a star (0.54~M$_\odot$) calculated using the Initial Mass
Function (IMF) 
\citep{Kro+93,Cha03}.  This results in 1.4$\times$10$^{11}$ stars.  
Applying the incidence of binaries, 
which is observed to be about 60\% of all systems investigated \citep{DM91}, 
to the 
number of stars in the Milky Way, the number of binary stars is 
determined to be 5.2$\times$10$^{10}$ 
(NB this is not just the number of stars times 
0.60, since each binary system contributes two stars).  

\item {\it Determination of the number of primaries with the
right mass and calculation of the mean primary's mass of our sample of binaries.}
Some stars have masses that are too low ($<$1.05 M$_\odot$; \citet{Por+98}) to evolve
off the main sequence in the average age of the galaxy 
(assumed to be 8.5 Gyr: \citet{LC00,Zoc+03}).
Stars with masses too high ($>$10M$_\odot$; \citet{Ibe95})
will never contribute to the PN population, but instead will undergo a
supernova explosion.
Using the IMF \citep{Kro01,Cha03}, we determined the number of binary systems
with primaries between
minimum and maximum mass limits (8.4\% of all binary systems in the Galaxy,
i.e., 4.4$\times$10$^9$ systems). This population of binaries is then
represented by
the mean mass of the primaries (1.78~M$_\odot$), calculated integrating the IMF
between our minimum and maximum mass values.

In the future we will not represent the entire population of binaries
by the mean mass of the primaries, but we will split the populations of primaries into
mass bins, each represented by a mean mass and we will
follow the evolution of each group. Also, the stellar lifetimes depend on the
metallicity of the star.
For further refinements of our calculation, we will
consider the Galaxy as continually star-forming and use the star formation
history as well as the age-metallicity relation to follow the
lives of stellar mass bins.

\item {\it Determination of the total number of binaries with the right
orbital separation}.
We assume that all post-CE binaries, where the CE happened on
the RGB, will not produce a PN (a $<$ 100 R$_\odot$).
Therefore, we are only concerned with
the binaries that have the orbital separation in a range such that the companion
will be engulfed as the primary expands during its ascent of the AGB.
This implies separations in the range 100
R$_\odot$ $<$ a $<$ 500 R$_\odot$ \citep{Dor+93}.  From the period
distribution of binaries \citep{DM91},
the number of systems within this range was
determined (12.1\% of all the binaries previously counted, i.e.,
5.3$\times$10$^8$ systems).
We currently assume that when a primary fills its Roche-Lobe
at the bottom of the AGB
the mass transfer is always unstable and a CE will always ensue. This is a
good assumption since the primary's envelope will quickly become convective.

\item {\it Binaries with the right mass ratio}. The companion must have sufficient
gravitational energy to unbind the envelope of the primary during the CE.
Therefore, the companion-to-primary mass ratio has to be large enough.
We took M$_2$/M$_1$=q $>$ 0.15 \citep{DeM+03} (see also below).
By investigating the observed distribution of mass
ratios \citep{DM91}, the fraction of binaries that will eject the AGB CE can be
determined (80.3\% of all systems previously counted, i.e., 4.3$\times$10$^8$
systems).

However, the mass ratio distribution \citep{DM91} is for main sequence binaries only;
q$>$0.15 however, applies to ratios of the companion to the mass of the
primary on the AGB, which is smaller than it was on the main sequence.
Thus, in future
calculations, we will have to account for the mass-loss rates up through to
the AGB phase \citep{Hur+00}, and derive the mass ratio at the time of CE from that
of main sequence stars.

It is also fundamental that the ejection of the envelope occurs when
the central core is luminous enough to sufficiently photo-ionize the ejected 
material and produce
a visible PN. If the initial binary separation is too small and the CE 
ejection happens
when the primary is at the very bottom of the AGB, 
the post-CE primary will have too low luminosity and will never
have enough hard photons to ionize the surrounding matter sufficiently for a PN
to be detected. We will account for this effect in our refined calculation.

\end{itemize}

Combining the previous steps yields the absolute number of galactic PN
formed via a CE ejection which harbor close binary central stars. If we assume
an average PN lifetime of 2.0$\times$10$^4$ years and an average stellar lifetime of 1.3$\times$10$^9$
years calculated from the average primary mass (1.78~M$_\odot$), 
then the total number of post-CE PN currently in our
galaxy is 6600 within a factor of 3 uncertainty. 

This number
should be compared to the actual number of galactic PN. The number of observed
PN in the Galaxy is $\sim$3000 \citep{Par+03}, but this
is a lower limit because of the large extinction on the galactic plane.
Rather than trying to estimate the total number of galactic PN extrapolating
their numbers by accounting for extinction,
a better estimate of the total number of galactic PN can be derived by counting
PN in external galaxies with similar morphology and mass. This number is
7200$\pm$1800 \citep{Pei90}, not dissimilar to the predicted number of PN
from CE ejections.
 
Using the calculation above we can also predict the number of PN in the galaxy if
both single and binary stars make a PN after ascending the AGB. This is
113\,000 within a factor of 2, hence much larger than the observationally-deduced number,
even allowing for the considerable uncertainty.
This means that only 5.6\% of stars (single or in binaries)
capable of ascending the AGB produce a PN via a CE ejection.

This fraction can be compared with the results of
\citet{Han+95}. \citet{Han+95}
determined that 13\% {\it of all binaries} result in a CE ejection from
the primary expansion on the RGB {\it and} AGB (where the secondary is a main
sequence stars). Only 0.3-0.5 of these binaries are post-CE {\it on the AGB}, 
according to their models. We must further multiply this fraction by 60\%,
the total binary fraction on the main sequence. The fraction of suitable
stars that produce a PN via a CE interaction on the AGB is therefore 2.5-4\%,
not dissimilar from our estimate. A detailed comparison between 
our calculation and the population synthesis model of \citet{Han+95} will
be presented in Moe \& De Marco (in prep.).

\section{Common envelope simulations}

\citet{DeM+03} carried out a series of 3-dimensional hydrodynamic simulations 
using the code and method of \citet{San+98}. Two of these simulations (Fig.~1)
were repeated at higher resolution and with a larger box size. These
simulate the effect of a 0.1-~M$_\odot$ companion entering CE with a
1.25-~M$_\odot$ primary at the bottom of the AGB or a 1.04-~M$_\odot$
primary at the top of the AGB (both these stellar models descend
from a 1.5-~M$_\odot$ main sequence primary). 

The most important result to come from these simulation is the extreme
sensitivity of the outcome of the CE interaction to the stellar and system
parameters: at the bottom of the AGB the more massive envelope with a smaller
radius is practically impervious to the penetration of the companion and
no envelope ejection occurs. The same small companion does eject the envelope
at the top of the AGB since by then the primary envelope is less massive and
more extended. The result is a short period binary with a period of about one
month. The primary-to-secondary mass ratio to eject the envelope is therefore
a sensitive function of when the secondary penetrates
the primary. Our choice of a value of 0.15 therefore leads to a lower limit to
the number of post CE binary central stars since if CE is entered
at the top o the AGB it appears that a ratio as low as 0.1 will suffice to
eject the envelope.

\begin{figure}
\includegraphics[height=.3\textheight,width=.45\textwidth]{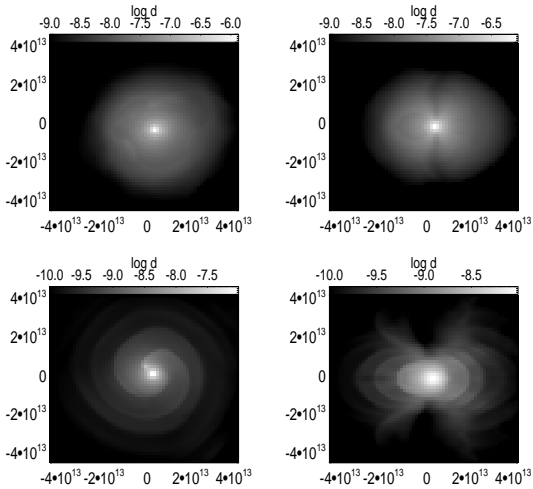}
  \caption{A greyscale of the density on the
orbital (left column) and perpendicular (right column) planes
cutting through our bottom (top row) and top (bottom row) of the AGB 
simulations at the end of the runs. The box size is marked on the sides
in cm. The greyscale intensity is labeled on the top in log scale, where
the units are in gr~cm$^{-3}$.}
\end{figure}

\section{Summary}

\begin{itemize}
\item There is a strong indication that the central star of PN close binary fraction might be much higher
than the 10\% believed thus far and possibly as high as 90-100\%. If confirmed this would establish
that PN are fundamentally a binary interaction byproduct.
\item Based on a population synthesis toy model it is not inconceivable that all of the PN present in
the Galaxy today derive from CE interactions that lead to the ejection of the envelope and leave behind
a short-period binary. 
\item If all of the single and binary post-AGB stars within a certain mass range produced PN
we should observe about 5-10 times more PN in the galaxy today.
\item To determine the CE parameters which are key in any population synthesis calculation we need
more theoretical knowledge of the interaction. From our calculation we can already say that
CE interactions can lead to a binary for secondary-to-primary mass ratios as small as 0.1.
\end{itemize}


\begin{theacknowledgments}
OD would like to acknowledge Janet Jeppson Asimov for financial support
and Howard Bond for delivering her talk in her stead.
\end{theacknowledgments}

\bibliography{demarco}

\end{document}